\documentclass[aps,prb,twocolumn,floatfix,showpacs,amsmath,amssymb]{revtex4}
\usepackage{amssymb}
\usepackage{amsmath}

\usepackage{graphicx}
\usepackage[hypertex]{hyperref}
\begin{document}

\title{Pressure tuning of Fermi surface topology of optimally doped BaFe$_1$$_.$$_9$Ni$_0$$_.$$_1$As$_2$}

\author{Xiao-Jia Chen,$^{1,2,3}$ Feng-Jiang Jia,$^{3}$ Jian-Bo Zhang,$^{3}$ Zhen-Xing Qin,$^{3}$ Ling-Yun Tang,$^{3}$ Qian Tao,$^{4}$ Zhu-An Xu,$^{4}$ Jing Liu,$^{5}$ Viktor V. Struzhkin,$^{1}$ Ronald E. Cohen,$^{1}$ and Ho-kwang Mao$^{1,2}$}
\affiliation{
$^{1}$Geophysical Laboratory, Carnegie Institution of Washington, Washington, DC 20015, USA\\
$^{2}$Center for High Pressure Science and Technology Advanced Research, Shanghai 201203, China\\
$^{3}$Department of Physics, South China University of Technology, Guangzhou 510640, China\\
$^{4}$Department of Physics, Zhejiang University, Hangzhou 310027, China\\
$^{5}$Institute of High Energy Physics, Chinese Academy of Sciences, Beijing 100190, China
}

\date{\today}

\begin{abstract}
The superconducting, transport, and structural properties of optimally electron-doped BaFe$_{1.9}$Ni$_{0.1}$As$_{2}$ are investigated by combining the electrical resistance and synchrotron X-ray diffraction measurements at high pressures. The superconducting transition temperature of this system is found to decrease in a similar way of the axial ratio of $c/a$ with increasing pressure but vanishing at a critical pressure of 7.5 GPa where $c/a$ has a dip and an isostructural transformation from a tetragonal to a collapsed tetragonal phase takes place. The resistance is found to obey a linear temperature dependence, evidencing the antiferromagnetic spin-fluctuations transport mechanism. The pressure effects are interpreted within the framework of pressure-induced Fermi surface topology modification in which pressure suppresses both the quasiparticle effective mass and the strength of the antiferromagnetic spin fluctuations leading to the reduction of superconductivity, accordingly. The absent superconductivity in the collapsed tetragonal phase is suggested to result from the complete suppression of the antiferromagnetic spin fluctuations.
\end{abstract}

\pacs{74.70.Xa, 74.62.Fj, 74.25.F-, 61.50.Ks}

\maketitle

\section{INTRODUCTION}

The recently discovered iron-based superconductors have been attracting great interests because of their high transition temperatures $T_{c}$'s  and the resemblance with the well-studied cuprates. Among all the iron pnictides, the Ba-based 122 family is so far the most studied iron-based system because of the availability of the high-purity single crystal samples.\cite{canf} This family shares the common phase diagram as other iron arsenides. For example, the undoped parent compound BaFe$_2$As$_2$ is magnetically ordered in a spin density wave (SDW) state. With cooling, this material undergoes both structural and magnetic transitions at about 140 K [\onlinecite{huang}]. At ambient pressure, superconductivity emerges upon doping by electrons,\cite{ljli} holes,\cite{mrot} or even isovalent elements\cite{shis,dhak} when the SDW instablity is suppressed. The maximum $T_{c}$'s are achieved close to the optimal concentration where the SDW state is completely suppressed. As a clean tuning parameter of structural and physical properties without introducing any disorder, the role of pressure in inducting superconductivity was particularly highlighted in the 122 families. All the 122 parent compounds become superconductors only under the application of pressure.\cite{hlee,milt,alir} This indicates that subtle change or contraction of the structure can switch on superconductivity from the neighboring antiferromagnetic (AF) ground state. Extensive studies on various Ba- [\onlinecite{alir,mst,fuka,ahil,hard,ishi,kahil,colo,mani,yama,klin,wuho,dunc,ecol,goh,drot,mitt,arse,kkim}], Sr- [\onlinecite{alir,mst,colo,gooc,igaw,koteg,walt}], Ca- [\onlinecite{hlee,milt,kasa}], and Eu-based [\onlinecite{tera,oya,lsun,kuri}] 122 families have established that the 122 compounds have a similar phase diagram with pressure as doping. Neutron diffraction measurements\cite{kimb} on the parent compound BaFe$_2$As$_2$ revealed many similarities between structural distortions under pressure and chemical doping. Quantum oscillation measurements\cite{graf} reported that the quasiparticle effective masses $m^{\star}$ in this parent in reality increases with increasing pressure, directly pointing to the importance of enhanced electronic correction for superconductivity.

Understanding pressure effect on superconductivity is important for synthesizing, designing, and discovering materials with higher $T_{c}$'s at ambient pressure by lattice modification in a way to simulate the pressure environment. The most successful example along this direction is the discovery of higher $T_{c}$'s in SmFeAsO$_{1-x}$F$_{x}$ through the chemical pressure.\cite{chen2} While the mechanism of pressure in inducing or tuning superconductivity remains unclear and is still a challenge, the optimally doped compounds are believed to be a good candidate for investigating such an effect due to the absence of the interplay and competition among the structural, magnetic, and superconducting orders. These compounds thus offer an ideal laboratory for elucidating the pressure effect on superconductivity with a maximum $T_{c}$ over the whole doping regime. There have been a lot of high-pressure measurements of electrical and magnetic transport properties of optimally doped Ba-based compounds Ba$_{1-x}$A$_{x}$(Fe$_{1-y}$B$_{y}$)$_{2}$(As$_{1-z}$C$_{z}$)$_{2}$ (A=K [\onlinecite{mst}]; B=Ru [\onlinecite{kkim}], Co [\onlinecite{ahil,kahil,ecol,drot,mitt,arse}]; C=P [\onlinecite{klin}]). However, these measurements were limited to lower pressures. The interesting physical properties in nonsuperconducting region at higher pressures have not been uncovered yet. No structural information has ever been reported for these optimally doped compounds. A comprehensive study on structural, transport, and superconducting properties of an optimally doped Ba-based 122 system at high pressures is therefore highly desirable.

In this paper, the above mentioned issues are addressed by combining the electrical resistance and synchrotron X-ray diffraction measurements on an optimally doped BaFe$_1$$_.$$_9$Ni$_0$$_.$$_1$As$_2$. We demonstrate that pressure-induced modification of Fermi surface topology accounts for the pressure effects on structural transformation and superconductivity evolution. Our results suggest that AF spin fluctuations should be the major player of the electrical transport, structural and superconducting properties in iron arsenic superconductors.

\section{EXPERIMENTAL DETAILS}

Single crystal sample BaFe$_{1.9}$Ni$_{0.1}$As$_{2}$ was grown by the self-flux method with the high-purity elements Ba, Fe, Ni using FeAs as the flux, detailed procedures of synthesizing the samples were reported previously.\cite{ljli} The electrical resistance measurements were performed by a standard four-probe method. Pressure is generated and changed to the sample by using a diamond anvil cell (DAC) with a pair of 500 $\mu$m culet size anvils. The principle is based on pushing two parallel aligned diamonds towards each other using a lever-arm system. A stainless steel gasket was preindented to 35 $\mu$m thick, and then a hole with a diameter of 400 $\mu$m was drilled in the center of the gasket. Cubic BN powders were put into the gasket hole and on the surface of one side of the gasket. After preindenting once more, an insulating environment was created for loading sample and putting electrical leads. Four electrical leads were then placed on such an insulating layer and a BaFe$_{1.9}$Ni$_{0.1}$As$_{2}$ crystal with dimensions of $60\times60\times20$ $\mu$m$^{3}$ was put on the tips of these electric leads. Ruby chips were put near the crystal and NaCl powders were dropped surrounding the crystal to serve as pressure transmitting medium. The pressure was monitored by the ruby fluorescence.\cite{mao}

High-pressure powder diffraction was carried out with the angle-dispersive X-ray diffraction experiments at the Beijing Synchrotron Radiation Facility (BSRF). A monochromatic X-ray beam with a wavelength of 0.6199 ${\AA}$ was used. A symmetric DAC with a pair of 300 $\mu$m culet size anvils was used to generate pressure. A stainless steel gasket preindented to 35 microns thick with a 120 $\mu$m diameter hole was used as the sample chamber. A small piece of sample pellet grounded from single crystals with a ruby ball was loaded in the gasket hole. Neon was used as pressure-transmitting medium to ensure better hydrostatic pressure condition. The pressures were also monitored by the ruby fluorescence shifts.\cite{mao}

\section{RESULTS AND DISCUSSION}

\subsection{Pressure effect on superconductivity}

Figure 1 shows the temperature dependence of the electrical resistance $R$ of BaFe$_1$$_.$$_9$Ni$_0$$_.$$_1$As$_2$ at pressure of 0.9 GPa and 11.5 GPa, respectively. The superconducting transition was signalled by the resistance drop. Here \emph{T$_c$} is determined from the intersection of the two extrapolated lines; one is drawn through the resistance curve in the normal state just above\emph{ T$_c$}, and the other is drawn through the steepest part of the resistance curve in the superconducting state. The \emph{T$_c$} value of this sample is 20.5 K at ambient pressure.\cite{ljli} We find that the \emph{T$_c$} is basically unchanged at pressure of 0.9 GPa. At 11.5 GPa, we were not able to detect superconducting transition drown to 10 K due to the limit of the experimental condition. Therefore, additional low-temperature electrical resistance experiments below 10 K are deserved to precisely pinpoint the disappearance of superconductivity. In this electrical resistance measurement, zero resistance was not observed even after a resistance drop. The failure for observing zero resistance is probably caused by a minor resistance in the contact between sample and electrodes or the contact between the electrodes and covered insulating layer on the gasket or a non-hydrostatic pressure condition created by the use of the solid pressure-transmitting medium.

\begin{figure}[tbp]
\includegraphics[width=\columnwidth]{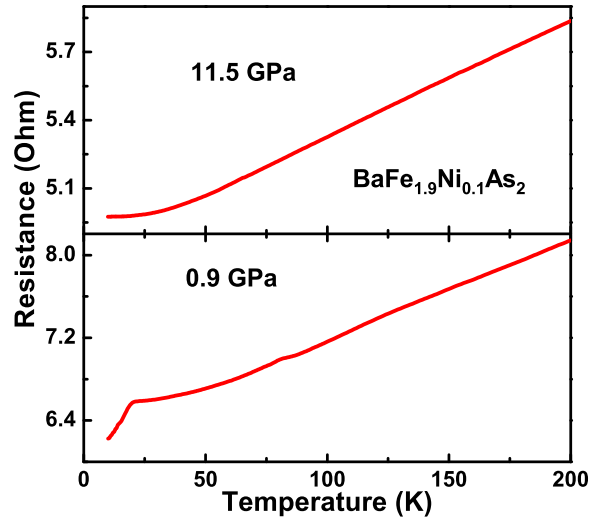}
\label{xijie}
\caption{ \label{model} (Color online)
The temperature dependence of the electrical resistance of BaFe$_1$$_.$$_9$Ni$_0$$_.$$_1$As$_2$ at 0.9 and 11.5 GPa,
respectively.}
\end{figure}

Figure 2 shows representative electrical resistance ratios \emph{R(T)/R(50 K)} of BaFe$_{1.9}$Ni$_{0.1}$As$_{2}$ single crystal as a function of temperature at different pressures. For pressures less than 2 GPa, the \emph{R(T)/R(50K)}-\emph{T} curves of this crystal are barely changed. With increasing pressure, the curves shift to the lower temperatures. Above 6.5 GPa, superconductivity can not be detected in our equipment capability down to 10 K, indicating superconductivity below 10 K or the absence of superconductivity. Although the zero-resistance for superconducting state has not been achieved for all pressures investigated, the superconducting transition is recovered after decompressing the sample to 3.3 GPa from the highest pressure of 16.5 GPa. This demonstrates that the failure of the detection of zero resistance at superconducting state does not affect the determination of the $T_{c}$ variation with pressure.

Figure 3 shows the pressure dependence of the determined \emph{T$_c$} in optimally doped BaFe$_1$$_.$$_9$Ni$_0$$_.$$_1$As$_2$ from the electrical resistance measurements. As can be seen, \emph{T$_c$} remains almost the initial value of 20 K at ambient condition below 2 GPa. With increasing pressure, \emph{T$_c$} suddenly drops. At applied pressure of 5.2 GPa, \emph{T$_c$} reaches 11.5 K. Upon compression from 6.5 GPa on, superconductivity if existed must be below 10 K. The arrow indicates that the \emph{T$_c$} is either below 10 K if superconductivity occurs or superconductivity completely disappears.

\begin{figure}[tbp]
\includegraphics[width=\columnwidth]{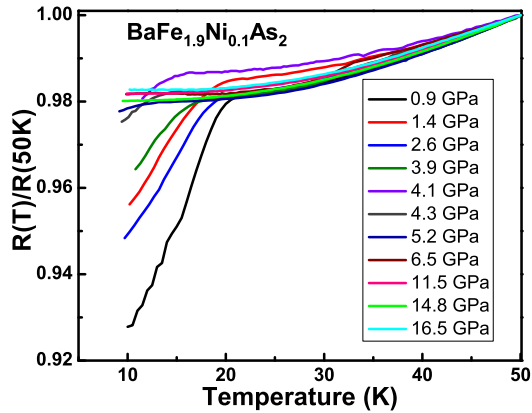}
\label{shuju}
\caption{(Color online) Representative electrical resistance ratio \emph{R(T)}/\emph{R}\emph{(50 K)} of
BaFe$_1$$_.$$_9$Ni$_0$$_.$$_1$As$_2$ as a function of temperature at different pressures.}
\end{figure}

We found that the measured $T_{c}$ data points at pressures up to 5.3 GPa fall well into the fitting curve by using the third order polynomial equation
\begin{equation}
\emph{T$_c$}=T_{c}(0)+\alpha\emph{P}+\beta\emph{P$^2$}+\gamma\emph{P$^3$} ~~\nonumber ,
\end{equation}
with the zero-pressure $T_{c}$(0)=20.18$\pm$0.77 K, $\alpha=-0.52\pm 1.31$ K/GPa, $\beta=-0.04\pm 0.57$ K/GPa$^{2}$, and $\gamma=-0.33\pm0.07$ K/GPa$^{3}$. When extending this tendency down to the absolute zero temperature, we obtained a pressure of 7.5 GPa at which superconductivity is believed to disappear. The solid line represents the polynomial fit to experimental \emph{T$_c$}\emph{(P)} data points. The difference between the observed resistance curve and fitting curve is quite reasonable. We thus determined the boundary between the superconducting and non-superconducting regions of this optimally doped material. After decompression from highest pressure to 3.3 GPa, the \emph{T$_c$} value is slightly hysteretic but still situates well on the \emph{T$_c$}-\emph{P} curve. It is suggested that the pressure effect on \emph{T$_c$} in this optimally doped BaFe$_{1.9}$Ni$_{0.1}$As$_2$ is reversible.

The obtained $T_{c}$ reduction in the optimally Ni-doped BaFe$_{2}$As$_{2}$ is similar to those in other Ba-based optimally doped 122 compounds.\cite{mst,kkim,ahil,kahil,ecol,drot,arse,klin} The detected pressure boundary between the superconducting and nonsuperconducting regime is helpful for uncovering the unknown behavior in nonsuperconducting region of these optimally doped 122 compounds. Among the studies on Ba-based 122 family, the substitution of Fe by Co is similar to the Ni case because both provide electron doping on the parent BaFe$_{2}$As$_{2}$. Compared to Fe$^2$$^+$ (3$d^6$) ion, Co$^2$$^+$ (3$d^7$) has one more 3$d$ electron, but Ni$^2$$^+$ (3$d^8$) has two more 3$d$ electrons. It is therefore expected that each Ni dopant induces two extra itinerant electrons, whereas each Co dopant induces only one extra itinerant electron. This offers a natural explanation for understanding the the optimal doping content of Ni equivalent to that of only about half of Co. Actually, both Ni-doped and Co-doped systems show maximum \emph{T$_c$ }at the same effective doping level. Therefore, the observed pressure effects in BaFe$_{1.9}$Ni$_{0.1}$As$_{2}$ should capture the basic high-pressure behavior in the sister BaFe$_{1.8}$Co$_{0.2}$As$_{2}$.

\begin{figure}[tbp]
\includegraphics[width=\columnwidth]{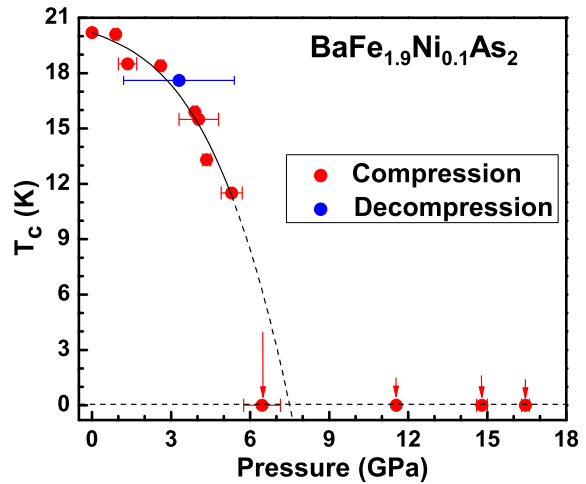}
\label{qushe}
\caption{(Color online) Pressure dependence of the superconducting transition temperature \emph{T$_c$} in optimally doped BaFe$_1$$_.$$_9$Ni$_0$$_.$$_1$As$_2$. The arrow indicates the \emph{T$_c$} below 10 K or the absence of superconductivity. The solid line is a fit of the data points to the third order polynomial function. }
\end{figure}

\subsection{Electrical transport}

In order to gain insight on the behavior of electrical resistance and superconductivity, we have fitted the temperature dependence of the resistances of BaFe$_1$$_.$$_9$Ni$_0$$_.$$_1$As$_2$ at various pressures using a simple power law $R=R_{0}+AT^{n}$, where the constant $R_0$ represents the contribution to the resistance from the electron-impurity scattering and the second term $AT^{n}$ accounts for the contribution to the resistance from the AF spin fluctuations with $n$=1 for non-Fermi liquid and the electron-electron interaction with $n$=2 for Fermi liquid. The fit to experimental data at 11.5 GPa is presented in Figure 4. It is seen that the experimental data fit yields $n$=1 in a board range of temperature from 200 K down to nearly 40 K. The resistances were found to generally obey the linear temperature dependence from 200 K down to 40 K for the pressures at which the sample is in the superconducting and non-superconducting regions. This electronic transport behavior is in good agreement with the observations in other optimally doped compound BaFe$_{1.8}$Co$_{0.2}$As$_{2}$ [\onlinecite{arse,ahil}].

The linear temperature dependence of the resistivity is the benchmark of strongly correlated electron phenomena in which the AF spin fluctuations are believed to play an essential role on superconductivity. Inelastic neutron scattering experiments on nearly optimally doped Ba$_{0.6}$K$_{0.4}$Fe$_{2}$As$_{2}$ [\onlinecite{chri}], BaFe$_{2-x}$Co$_{x}$As$_{2}$ ($x$=0.15,0.16) [\onlinecite{inos,lums}], and BaFe$_{1.9}$Ni$_{0.1}$As$_{2}$ [\onlinecite{chi,sli,zhao,nliu}] revealed the opening of a gap in the spin-fluctuation spectrum and the development of a resonant spin excitation at energies scaling well with $T_{c}$. These results highlight that the AF spin fluctuations are the most competitive driving force for the mechanism of both the electronic transport and superconductivity in optimally doped 122 compounds. A microscopic theory of superconductivity for iron arsenides has been developed with the emphasis the role of spin excitations.\cite{mazi,kors,maie}

\begin{figure}[tbp]
\includegraphics[width=\columnwidth]{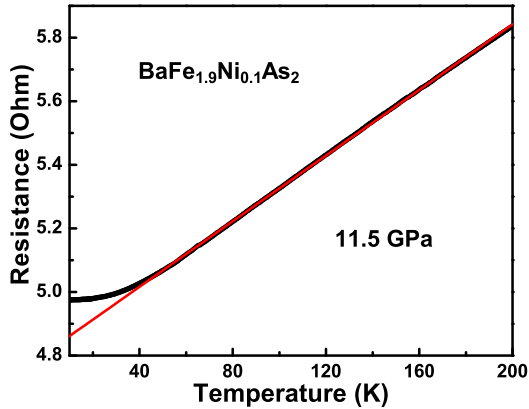}
 \label{fit}
 \caption{(Color online) The fit of the normal state resistance of BaFe$_1$$_.$$_9$Ni$_0$$_.$$_1$As$_2$ at 11.5 GPa using a simple power law $R$=$R_{0}$+$AT^{n}$. The red line represents the linear temperature behavior of the resistance.}
\end{figure}

\subsection{Structural properties}

High-pressure X-ray diffraction measurements were performed to understand the \emph{T$_c$ }-\emph{P} phase diagram of the optimally doped BaFe$_1$$_.$$_9$Ni$_0$$_.$$_1$As$_2$. Figure 5 shows the pressure dependence of the lattice parameters $a$ and $c$ of this material at room temperature. In the parent compound BaFe$_{2}$As$_{2}$ [\onlinecite{mitt,yang}], the lattice parameter $a$ always exhibits intermediate anomalous expansion with a $S$ shape at high pressures. The obtained $a$ and $c$ in this optimally doped material rather monotonically decrease with increasing pressure, yielding a monotonic decrease in the unit-cell volume $V$ without any observable kink or discontinuity (e.g. Fig. 6). As seen in Fig. 6, the axial ratio $c/a$ varies with pressure in a parabolic-like way with a dip at a critical pressure between 6.8 and 8.7 GPa. This critical pressure level is exactly consistent to that of 7.5 GPa where superconductivity disappears. The axial ratio $c/a$ of 3.22 at ambient conditions in this optimally doped compound is smaller than 3.3 in the parent compound.\cite{mitt,yang} It is indicated that the critical pressure for structural transformation in  BaFe$_{2}$As$_{2}$ should be higher than that in BaFe$_1$$_.$$_9$Ni$_0$$_.$$_1$As$_2$. This is evidenced from the structural measurements.\cite{mitt,yang}

\begin{figure}[tbp]
\begin{center}
\includegraphics[width=\columnwidth]{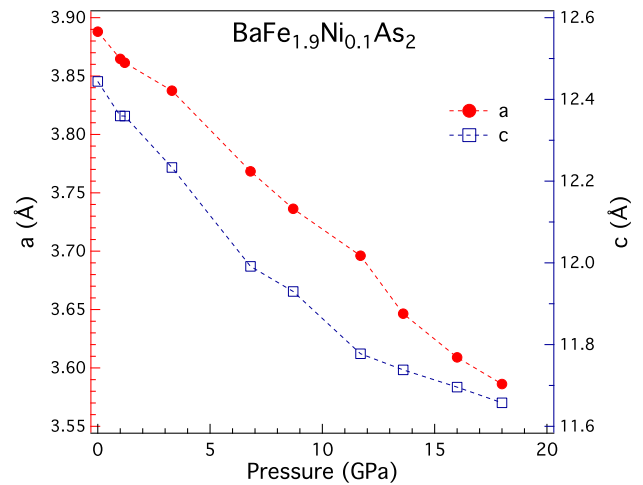}
\end{center}
\caption{(Color online) Pressure dependence of the lattice parameters $a$ and $c$ of optimally doped BaFe$_{1.9}$Ni$_{0.1}$As$_{2}$ up to 18 GPa and room temperature. }
\end{figure}

It has been found\cite{mitt,yang,uhoy} that superconductivity is absent in high-pressure collapsed tetragonal ($cT$) phase in other 122 systems. The sudden increase in the $c/a$ ratio at higher pressures beyond a critical pressure together with the absence of superconductivity indicates that the optimally doped BaFe$_{1.9}$Ni$_{0.1}$As$_2$ enters a $cT$ phase upon heavy compression. Recently, Sun $et$ $al.$ [\onlinecite{sun}] found that there are two superconducting regions in the 122*-type iron-based superconductors and the second region has a much higher the \emph{T$_c$} than the first one when passing a nonsuperconducting intermediate region. At the highest pressure of 16.5 GPa studied, there is no indication of the appearance of  superconductivity in the optimally doped BaFe$_{1.9}$Ni$_{0.1}$As$_2$.

\subsection{Fermi surface topology}

Now we explain the observed pressure effects on the superconducting, transport, and structural properties in optimally doped BaFe$_1$$_.$$_9$Ni$_0$$_.$$_1$As$_2$ as a whole. A comparison study\cite{bald} of doping and pressure effects on 122 compounds suggested that the mechanism for driving superconductivity becomes common only if the system in the overdoped regime where an increased itinerancy is responsible for $T_{c}$ reduction in both cases. It is therefore expected that our observations and explanations are also appropriate to this electron doped system in the overdoped regime.

\begin{figure}[tbp]
\begin{center}
\includegraphics[width=\columnwidth]{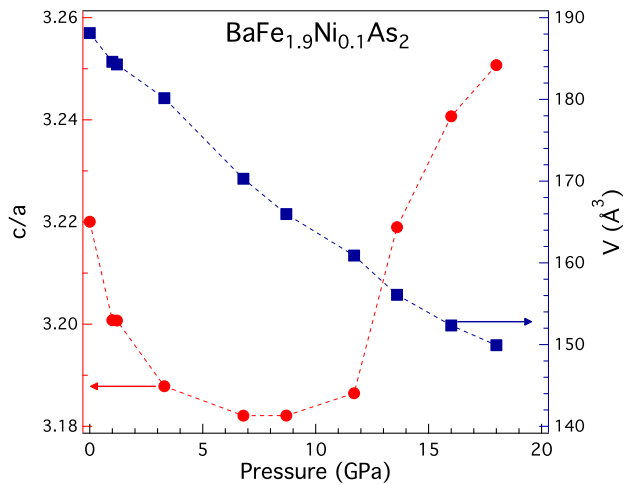}
\end{center}
\caption{(Color online) Pressure dependence of the axial ratio $c/a$ and unit-cell volume $V$ of optimally doped BaFe$_{1.9}$Ni$_{0.1}$As$_{2}$ up to 18 GPa and room temperature. }
\end{figure}

The Fermi surface consists of small hole and electron pockets in iron-pnicitide superconductors. In general, the application of pressure can modify the Fermi surface topological structure, leading to the changes of the density of states (primarily Fe 3$d$ contribution) $N(E_{F})$ at the Fermi energy and the carrier effective masses $m^{\star}$. Since $N(E_{F})$ generally correlates with $T_{c}$ in conventional BCS superconductors, it is natural to think whether the pressure-induced change of $N(E_{F})$ could account for the variation $T_{c}$ in iron-pnicitide superconductors.\cite{wei} Our observation of spin-fluctuations dominated quasiparticle transport suggests that scattering electrons by spin fluctuations from the electron to the hole sheets should become weaker with increasing pressure. This effect probably leads to an expansion of both sheet volumes and a reduction in $m^{*}$ accordingly. Recently, the AF spin fluctuations were indeed found to be substantially reduced with increasing pressure in the optimally doped and overdoped BaFe$_{2-x}$Ni$_{x}$As$_{2}$ ($x$=0.1,0.2) [\onlinecite{dong}]. Our obtained $T_{c}$ reduction with pressure in optimally doped BaFe$_1$$_.$$_9$Ni$_0$$_.$$_1$As$_2$ is thus suggested to result from the pressure-induced reduction of the AF spin fluctuations. Thus, the pressure-induced reduction of $N(E_{F})$ may not be a dominant factor in controlling $T_{c}$ behavior. This is indicated from the observations of the variation of $T_{c}$ with the isovalent element substitution of Fe by Ru or As by P in BaFe$_{2}$As$_{2}$ by a nearly unperturbed $N(E_{F})$ [\onlinecite{shis,dhak}] and the almost constant maximum $T_{c}$ in BaFe$_{2}$(As$_{1-x}$P$_{x}$)$_{2}$ at high pressures from the undoped to optimally doped region.\cite{klin}

The application of pressure on the optimally electron doped system brings about the expansion of the volumes of electron and hole Fermi surfaces (yielding a reduction in $m^{*}$) and the reduction of the strength of the AF spin fluctuations. Due to the similarity between the overdoping and pressure, doping should have the similar effect. Measurements of the de Haas-van Alphen effect\cite{shis} on BaFe$_{2}$(As$_{1-x}$P$_{x}$)$_{2}$ indeed revealed a strong reduction in $m^{*}$ when doping from the optimal level to the overdoped side. Nuclear magnetic resonance (NMR) measurements on this system confirmed that the AF spin fluctuations monotonically decreases when doping away from the optimal level.\cite{naka} Magnetic measurements on optimally doped BaFe$_{2}$(As$_{0.65}$P$_{0.35}$)$_{2}$ [\onlinecite{goh}] further suggested that the $T_{c}$ reduction with pressure is mainly due to the pressure-induced reduction of $m^{*}$ (or AF spin fluctuations) in the $ab$ plane. For electron doped 122 systems, a reduction in the strength of the AF spin fluctuations with doping in the overdoped regime has been found from both inelastic neutron scattering and NMR measurements.\cite{lums,inos,chi,sli,zhao,nliu,mwang,ning} A correlation between $T_{c}$ and the strength of the AF spin fluctuations has been well established.

A close relationship between the axial ratio $c/a$ and superconductivity has been found in the parent 122 compounds.\cite{yang,uhoy} It is indicated that the three-dimensional feature of the structure in these compounds is in favor of superconductivity under pressure. The observations of the consistent evolution of both the $T_{c}$ and $c/a$ in the tetragonal phase in optimally doped BaFe$_{1.9}$Ni$_{0.1}$As$_{2}$ offer a strong support for this relationship. It is clear that the reduction of superconduction under pressure is due to the weakness of the three-dimensional structural feature. In fact, the three-dimensional nature of the superconducting gap and its sensitivity to the AF ordering wave vectors $Q$ along the $c$ axis were reported from inelastic neutron scattering experiments on the same optimally doped material.\cite{chi,sli,zhao,nliu} Specifically, Wang $et$ $al.$ [\onlinecite{mwang}] found that the anisotropic spin gaps at the two wave vectors $Q=$(0.5,0.5,0) and (0.5,0.5,1) decrease and spin excitations gradually transform into quasi-two-dimensional spin excitations with increasing electron-doping in BaFe$_{2-x}$B$_{x}$As$_{2}$ (B=Ni,Co). As a result, both the $T_{c}$ and spin resonance energies decrease on the overdoped side. It is thus indicated that the axial ratio $c/a$ is a good measure of the strength of the AF spin fluctuations and thus the carrier effective mass in these 122 compounds.

At higher pressures, the 122 iron-based superconductors often undertake an isostructural transformation from a tetragonal to a $cT$ phase at a critical pressure where superconductivity disappears.\cite{mitt,yang,jia} For the optimally doped BaFe$_{1.9}$Ni$_{0.1}$As$_{2}$ compound studied, the critical pressure is determined to be 7.5 GPa after which the material is no longer superconducting and the $c/a$ ratio climbs. It is therefore indicated that the isostructural phase transition occurs at 7.5 GPa. At higher pressures, the material enters a $cT$ phase. This collapse was suggested to be the onset of $4p_{z}$ interlayer bond formation.\cite{kasi} Recently, the suppression of AF spin fluctuations in the $cT$ phase was observed from the high-pressure inelastic neutron scattering measurements on CaFe$_{2}$As$_{2}$ [\onlinecite{pratt}]. The absence of superconductivity in this $cT$ phase was found to be a result of the suppression or absence of both static AF order and dynamic spin fluctuations. Thus, the pressure-induced suppression of the AF spin fluctuations must account for the absence of superconductivity in optimally doped BaFe$_{1.9}$Ni$_{0.1}$As$_{2}$ at higher pressures above 7.5 GPa. Pressure-tuned change of Fermi surface topology offers a natural explanation on the pressure effects on structural and superconducting properties of optimally doped BaFe$_{1.9}$Ni$_{0.1}$As$_{2}$.

The observed $c/a$ evolution with pressure indicates that the ratio $c/a$ could be the sensor of Fermi surface topology in the optimally doped BaFe$_{1.9}$Ni$_{0.1}$As$_{2}$. The decrease in $c/a$ with pressure supports the theoretical prediction that three dimensionality of Fermi surface leads to a more gradual decrease of $T_{c}$ [\onlinecite{fern}]. Recent angle-resolved photoemission spectroscopy studies\cite{liu} on Ba(Fe$_{1-x}$Co$_{x}$)$_{2}$As$_{2}$ revealed that the $Z$ pocket undergoes a drastic topological change from hole-like to electron-like at roughly a critical doping level where superconductivity vanishes. Similar to the observed $c/a$ shape with pressure, the $Z$ pocket in Ba(Fe$_{1-x}$Co$_{x}$)$_{2}$As$_{2}$ was observed to behave parabolically in shape with electron doping. The size of this hole pocket shrinks and approaches to zero at the critical level, above which an electron pocket appears and increases in size with further doping. The critical doping level marks the Lifshitz transition. If the overdoping and pressure also affect superconductivity in a similar way, the Lifshitz transition should also take place at our obtained critical pressure in BaFe$_{1.9}$Ni$_{0.1}$As$_{2}$.

It should be noticed that the pressure-induced modification of Fermi surface topology is more complicated for the undoped and underdoped 122 compounds based on BaFe$_{2}$As$_{2}$ [\onlinecite{mst,alir,fuka,ahil,hard,ishi,kahil,colo,mani,yama,klin,wuho,dunc,ecol,goh,drot,mitt,arse,kkim}], SrFe$_{2}$As$_{2}$ [\onlinecite{mst,alir,gooc,igaw,colo,koteg,walt}], CaFe$_{2}$As$_{2}$ [\onlinecite{milt,hlee,kasa}], and EuFe$_{2}$As$_{2}$ [\onlinecite{tera,oya,lsun,kuri}] due to the competition of magnetic order and superconductivity, although there exist similarities between structural distortions under pressure and chemical doping.\cite{kimb} Application of pressure always brings about the suppression of magnetic order and the emergence of superconductivity for these compounds. In contrast to the reduction of the AF spin fluctuations with pressure or doping in the optimally doped level or overdoped region, the applied pressure usually leads to the enhancement of the AF spin fluctuations and carrier effective mass in the undoped or underdoped compound.\cite{imai,graf} Therefore, the underdoped and overdoped regions in Ba-based 122 system have different superconducting behaviors. A sharp peak of the zero-temperature penetration depth was observed at optimal composition in BaFe$_{2}$(As$_{1-x}$P$_{x}$)$_{2}$ [\onlinecite{khas}]. These results imply a possible crossover toward the Bose-Einstein condensate limit driven by quantum criticality.

\section{CONCLUSIONS}

In conclusion, we have investigated the superconducting, transport, and structural properties of optimally electron-doped BaFe$_{1.9}$Ni$_{0.1}$As$_{2}$ by combining the electrical resistance and synchrotron X-ray diffraction measurements. We found that $T_{c}$ remains its ambient-condition value at pressures below 2 GPa but it decreases with increasing pressure reaching 11.5 K at 5.2 GPa. Fitting the measured $T_{c}$ data points yields a critical pressure at around 7.5 GPa at which superconductivity is expected to disappear. Structural analysis shows that the axial ratio $c/a$ monotonically decreases with increasing pressure but reaching a minimum at around 7.5 GPa. The similar behavior of both the $T_{c}$ and $c/a$ implies that the $c/a$ ratio is a structural feature controlling superconductivity. The resistance in the temperature range between 40 and 200 K was found to obey a linear temperature dependence, evidencing the importance of the antiferromagnetic spin fluctuations in this optimally electron-doped compound. We proposed that pressure-induced Fermi surface topology accounts for all these observed effects. The application of pressure leads to the reduction of both the carrier effective mass and the strength of the antiferromagnetic spin flucturations and the suppression of superconductivity, accordingly. When pressure is applied above the critical pressure of 7.5 GPa, the material enters a collapsed tetragonal phase where the antiferromagnetic spin fluctuations is completely suppressed and the superconductivity is absent. These results suggest that antiferromagnetic spin fluctuations are the the major player of superconductivity and physical properties in iron-pnicitide superconductors.

\section{ACKNOWLEDGMENTS}

This work at Carnegie is supported by EFree, an Energy Frontier Research Center funded by DOE-BES under grant number DE-SC0001057. The sample synthesis and measurements in China are supported by China 973 Program (Grant No. 2011CBA00103), NSFC (Grant No. 11174247), and the Cultivation Fund of the Key Scientific and Technical Innovation Project Ministry of Education of China (No.708070). 4W2 beamline of BSRF is supported by CAS (Grant Nos. KJCX2-SW-N20 and KJCX2-SW-N03).


\begin{thebibliography}{2dspins}

\bibitem{canf} P. C. Canfield and S. L. Bud'ko, Annu. Rev. Condens. Matter Phys. {\bf 1}, 27 (2010).

\bibitem{huang} Q. Huang, Y. Qiu, W. Bao, M. A. Green, J. W. Lynn, Y. C. Gasparovic, T. Wu, G. Wu, and X. H. Chen, Phys. Rev. Lett. {\bf 101}, 257003 (2008).

\bibitem{ljli} L. J. Li, Y. K. Luo, Q. B. Wang, H. Chen, Z. Ren, Q. Tao, Y. K. Li, X. Lin, M. He, Z. W. Zhu, G. H. Cao, and Z. A. Xu, New J. Phys. \textbf{11}, 025008 (2009).

\bibitem{mrot} M. Rotter, M. Tegel, and D. Johrendt, Phys. Rev. Lett. \textbf{101}, 107006 (2008).

\bibitem{shis} H. Shishido, A. F. Bangura, A. I. Coldea, S. Tonegawa, K. Hashimoto, S. Kasahara, P. M. C. Rourke, H. Ikeda, T. Terashima, R. Settai, Y. Onuki, D. Vignolles, C. Proust, B. Vignolle, A. McCollam, Y. Matsuda, T. Shibauchi, and A. Carrington, Phys. Rev. Lett. {\bf 104}, 057008 (2010).

\bibitem{dhak} R. S. Dhaka, C. Liu, R. M. Fernandes, R. Jiang, C. P. Strehlow, T. Kondo, A. Thaler, J. Schmalian, S. L. Bud'ko, P. C. Canfield, and A. Kaminski, Phys. Rev. Lett. {\bf 107}, 267002 (2011).

\bibitem{hlee} H. Lee, E. Park, T. Park, V. A. Sidorov, F. Ronning, E. D. Bauer, and J. D. Thompson, Phys. Rev. B {\bf 80}, 024519 (2009).

\bibitem{milt} M. S. Torikachvili, S. L. Bud'ko, N. Ni, and P. C. Canfield, Phys. Rev. Lett. {\bf 101}, 057006 (2008).

\bibitem{alir} P. L. Alireza, Y. T. Chris Ko, J. Gillett, C. M. Petrone, J. M. Cole, G. G. Lonzarich, and S. E. Sebastian, J. Phys.: Condens. Matter \textbf{21}, 012208 (2009).

\bibitem{mst} M. S. Torikachvili, S. L. Bud'ko, N. Ni, and P. C. Canfield, Phys. Rev. B {\bf 78}, 104527 (2008).

\bibitem{fuka} H. Fukazawa, N. Takeshita, T. Yamazaki, K. Kondo, K. Hirayama, Y. Kohori, K. Miyazawa, H. Kito, H. Eisaki, and A. Iyo, J. Phys. Soc. Jpn. {\bf 77}, 105004 (2008).

\bibitem{ahil} K. Ahilan, J. Balasubramaniam, F. L. Ning, T. Imai, A. S. Sefat, R. Jin, M. A. McGuire, B. C. Sales, and D. Mandrus, J. Phys.: Condens. Matter {\bf 20}, 472201 (2008).

\bibitem{hard} F. Hardy, P. Adelmann, T. Wolf, H. v. L\"{o}hneysen, and C. Meingast, Phys. Rev. Lett. {\bf 102}, 187004 (2009).

\bibitem{ishi} F. Ishikawa, N. Eguchi, M. Kodama, K. Fujimaki, M. Einaga, A. Ohmura, A. Nakayama, A. Mitsuda, and Y. Yamada, Phys. Rev. B {\bf 79}, 172506 (2009).

\bibitem{kahil} K. Ahilan, F. L. Ning, T. Imai, A. S. Sefat, M. A. McGuire, B. C. Sales, and D. Mandrus, Phys. Rev. B {\bf 79}, 214520 (2009).

\bibitem{colo} E. Colombier, S. L. Bud'ko, N. Ni, and P. C. Canfield, Phys. Rev. B {\bf 79}, 224518 (2009).

\bibitem{mani} A. Mani, N. Ghosh, S. Paulraj, A. Bharathi, and C. S. Sundar, EPL {\bf 87}, 17004 (2009).

\bibitem{yama} T. Yamazaki, N. Takeshita, R. Kobayashi, H. Fukazawa, Y. Kohori, K. Kihou, C.-H. Lee, H. Kito, A. Iyo, and H. Eisaki, Phys. Rev. B {\bf 81}, 224511 (2010).

\bibitem{klin} L. E. Klintberg, S. K. Goh, S. Kasahara, Y. Nakai, K. Ishida, M. Sutherland, T. Shibauchi, Y. Matsuda, and T. Terashima, J. Phys. Soc. Jpn. {\bf 79}, 123706 (2010).

\bibitem{wuho} W. Uhoya, A. Stemshorn, G. Tsoi, Y. K. Vohra, A. S. Sefat, B. C. Sales, K. M. Hope, and S. T. Weir, Phys. Rev. B {\bf 82}, 144118 (2010).

\bibitem{dunc} W. J. Duncan, O. P. Welzel, C. Harrison, X. F. Wang, X. H. Chen, F. M. Grosche, and P. G. Niklowitz, J. Phys.: Condens. Matter {\bf 22}, 052201 (2010).

\bibitem{ecol} E. Colombier, M. S. Torikachvili, N. Ni, A. Thaler, S. L. Bud'ko, and P. C. Canfield, Supercond. Sci. Technol. {\bf 23}, 054003 (2010).

\bibitem{goh} S. K. Goh, Y. Nakai, K. Ishida, L. E. Klintberg, Y. Ihara, S. Kasahara, T. Shibauchi, Y. Matsuda, and T. Terashima, Phys. Rev. B 82, 094502 (2010).

\bibitem{drot} S. Drotziger, P. Schweiss, K. Grube, T. Wolf, P. Adelmann, C. Meingast, and H. v. L\"{o}hneysen, J. Phys. Soc. Jpn. {\bf 79}, 124705 (2010).

\bibitem{mitt} R. Mittal, S. K. Mishra, S. L. Chaplot, S. V. Ovsyannikov, E. Greenberg, D. M. Trots, L. Dubrovinsky, Y. Su, T. Brueckel, S. Matsuishi, H. Hosono, and G. Garbarino, Phys. Rev. B {\bf 83}, 054503 (2011).

\bibitem{arse} S. Arsenijevic, R. Gaal, A. S. Sefat, M. A. McGuire, B. C. Sales, D. Mandrus, and L. Forro, Phys. Rev. B {\bf 84}, 075148 (2011).

\bibitem{kkim} S. K. Kim, M. S. Torikachvili, E. Colombier, A. Thaler, S. L. Bud'ko, and P. C. Canfield, Phys. Rev. B {\bf 84}, 134525 (2011).

\bibitem{gooc} M. Gooch, B. Lv, B. Lorenz, A. M. Guloy, and C.-W. Chu, Phys. Rev. B  \textbf{78}, 180508 (2008).

\bibitem{igaw} K. Igawa, H. Okada, H. Takahashi, S. Matsuishi, Y. Kamihara, M. Hirano, H. Hosono, K. Matsubayashi, and Y. Uwatoko, J. Phys. Soc. Jpn. {\bf 78}, 025001 (2009).

\bibitem{koteg} H. Kotegawa, T. Kawazoe, H. Sugawara, K. Murata, and H. Tou, J. Phys. Soc. Jpn. {\bf 78}, 083702 (2009).

\bibitem{walt} W. O. Uhoya, J. M. Montgomery, G. M. Tsoi, Y. K. Vohta, M. A. McGuire, A. S. Sefat, B. C. Sales, and S. T. Weir, J. Phys.: Condens. Matter {\bf 23}, 122201 (2011).

\bibitem{kasa} S. Kasahara, T. Shibauchi, K. Hashimoto, Y. Nakai, H. Ikeda, T. Terashima, and Y. Matsuda, Phys. Rev. B {\bf 83}, 060505(R) (2011).

\bibitem{tera} T. Terashima, M. Kimata, H. Satsukawa, A. Harada, K. Hazama, S. Uji, H. S. Suzuki, T. Matsumoto, and K. Murata, J. Phys. Soc. Jpn. {\bf 78}, 083701 (2009).

\bibitem{oya} W. O. Uhoya, G. M. Tsoi, Y. K. Vohta, M. A. McGuire, A. S. Sefat, B. C. Sales, D. Mandrus, and S. T. Weir, J. Phys.: Condens. Matter {\bf 22}, 292202 (2010).

\bibitem{lsun} L. L. Sun, J. Guo, G. F. Chen, X. H. Chen, X. L. Dong, W. Lu, C. Zhang, Z. Jiang, Y. Zou, S. Zhang, Y. Y. Huang, Q. Wu, X. Dai, Y. C. Li, J. Liu, and Z. X. Zhao, Phys. Rev. B {\bf 82}, 134509 (2010).

\bibitem{kuri} N. Kurita, M. Kimata, K. Kodama, A. Harada, M. Tomita, H. S. Suzuki, T. Matsumoto, K. Murata, S. Uji, and T. Terashima, Phys. Rev. B {\bf 83}, 214513 (2011).

\bibitem{kimb} S. A. J. Kimber, A. Kreyssig, Y.-Z. Zhang, H. O. Jeschke, R. Valentí, F. Yokaichiya, E. Colombier, J. Yan, T. C. Hansen, T. Chatterji, R. J. McQueeney, P. C. Canfield, A. I. Goldman, and D. N. Argyriou, Nat. Mater. {\bf 8}, 471 (2009).

\bibitem{graf} D. Graf, R. Stillwell, T. P. Murphy, J.-H. Park, E. C. Palm, P. Schlottmann, R. D. McDonald, J. G. Analytis, I. R. Fisher, and S. W. Tozer, Phys. Rev. B {\bf 85}, 134503 (2012).

\bibitem{chen2} X. H. Chen, T. Wu, R. H. Liu, H. Chen, and D. F. Fang, Nature (London) {\bf 453}, 761 (2008).

\bibitem{mao} H. K. Mao, J. Xu and P. M. Bell, J. Geophys. Res. \textbf{91}, 4673 (1986).

\bibitem{chri} A. D. Christianson, E. A. Goremychkin, R. Osborn, S. Rosenkranz, M. D. Lumsden, C. D. Malliakas, I. S. Todorov, H. Claus, D. Y. Chung, M. G. Kanatzidis, R. I. Bewley, and T. Guidi, Nature (London) {\bf 456}, 930 (2008).

\bibitem{inos} D. S. Inosov, J. T. Park, P. Bourges, D. L. Sun, Y. Sidis, A. Schneidewind, K. Hradil, D. Haug, C. T. Lin, B. Keimer, and V. Hinkov, Nat. Phys. {\bf 6}, 178 (2010).

\bibitem{lums} M. D. Lumsden, A. D. Christianson, D. Parshall, M. B. Stone, S. E. Nagler, G. J. MacDougall, H. A. Mook, K. Lokshin, T. Egami, D. L. Abernathy, E. A. Goremychkin, R. Osborn, M. A. McGuire, A. S. Sefat, R. Jin, B. C. Sales, and D. Mandrus, Phys. Rev. Lett. \textbf{102}, 107005 (2009).

\bibitem{chi} S. X. Chi, A. Schneidewind, J. Zhao, L. W. Harriger, L. J. Li, Y. K. Luo, G. H. Cao, Z. A. Xu, M. Loewenhaupt, J. P. Hu, and P. C. Dai, Phys. Rev. Lett. \textbf{102},107006 (2009).

\bibitem{sli} S. L. Li, Y. Chen, S. Chang, J. W. Lynn, L. J. Li, Y. K. Luo, G. H. Cao, Z. A. Xu, and P. C. Dai, Phys. Rev. B  \textbf{79}, 174527 (2009).

\bibitem{zhao} J. Zhao, L. P. Regnault,  C. L. Zhang,  M. Y. Wang, Z. C. Li,  F. Zhou,  Z. X. Zhao, C. Fang, J. P. Hu, and P. C. Dai, Phys. Rev. B  \textbf{81}, 180505(R) (2010).

\bibitem{nliu} M. S. Liu, L. W. Harriger, H. Q. Luo, M. Wang, R. A. Ewings, T. Guidi, H. Park, K. Haule, G. Kotliar, S. M. Hayden, and P. C. Dai, Nat. Phys. {\bf 8}, 376 (2012).

\bibitem{mazi} I. I. Mazin, D. J. Singh, M. D. Johannes, and M. H. Du, Phys. Rev. Lett. {\bf 101}, 057003 (2008).

\bibitem{kors} M. M. Korshunov and I. Eremin, Phys. Rev. B {\bf 78}, 140509(R) (2008).

\bibitem{maie} T. A. Maier, S. Graser, D. J. Scalapino, and P. Hirschfeld, Phys. Rev. B {\bf 79}, 134520 (2009).

\bibitem{yang} W. G. Yang, F. J. Jia, L. Y. Tang, L. J. Li, Z. A. Xu, and X. J. Chen, Phys. Rev. B (submitted).

\bibitem{uhoy} W. O. Uhoya, J. M. Montgomery, G. M. Tsoi, Y. K. Vohra, M. A. McGuire, A. S. Sefat, B. C. Sales, and S. T. Weir, J. Phys.: Condens. Matter {\bf 23}, 122201 (2011).

\bibitem{sun} L. L. Sun, X. J. Chen, J. Guo, P. W. Gao, Q.-Z. Huang, H. D. Wang, M. H. Fang, X. L. Chen, G. F. Chen, Q. Wu, C. Zhang, D. C. Gu, X. L. Dong, L. Wang, K. Yang, A. G. Li, X. Dai, H.-K. Mao and Z. X. Zhao, Nature (London) \textbf{483}, 67 (2012).

\bibitem{bald} L. Baldassarre, A. Perucchi, P. Postorino, S. Lupi, C. Marini, L. Malavasi, J. Jiang, J. D. Weiss, E. E. Hellstrom, I. Pallecchi, and P. Dore, Phys. Rev. B {\bf 85}, 174522 (2012).

\bibitem{wei} W. Yi, L. L. Sun, Z. A. Ren, W. Lu, X. L. Dong, H. J. Zhang, X. Dai, Z. Fang, Z. C. Li, G. C. Che, J. Yang, X. L. Shen, F. Zhou, and Z. X. Zhao, EPL {\bf 83}, 57002 (2008).

\bibitem{dong} X. D. Zhang, G. Z. Fan, C. L. Zhang, X. N. Jing, and J. L. Luo, Chin. Phys. Lett. {\bf 29}, 017401 (2012).

\bibitem{naka} Y. Nakai, T. Iye, S. Kitagawa, K. Ishida, H. Ikeda, S. Kasahara, H. Shishido, T. Shibauchi, Y. Matsuda, and T. Terashima, Phys. Rev. Lett. {\bf 105}, 107003 (2010).

\bibitem{ning} F. L. Ning, K. Ahilan, T. Imai, A. S. Sefat, M. A. McGuire, B. C. Sales, D. Mandrus, P. Cheng, B. Shen, and H.-H. Wen, Phys. Rev. Lett. {\bf 104}, 037001 (2010).

\bibitem{mwang} M. Y. Wang, H. Q. Luo, J. Zhao, C. L. Zhang,  M. Wang, K.Marty, S. X. Chi,  J. W. Lynn, A. Schneidewind,  S. L. Li, and P. C. Dai, Phys. Rev. B \textbf{81}, 174524 (2010).

\bibitem{jia} F. J. Jia, W. G. Yang, L. J. Li, Z. A. Xu, and X. J. Chen, Physica C {\bf 474}, 1 (2012).

\bibitem{kasi} D. Kasinathan, M. Schmitt, K. Koepernik, A. Ormeci, K. Meier, U. Schwarz, M. Hanfland, C. Geibel, Y. Grin, A. Leithe-Jasper, and H. Rosner, Phys. Rev. B {\bf 84}, 054509 (2011).

\bibitem{pratt} D. K. Pratt, Y. Zhao, S. A. J. Kimber, A. Hiess, D. N. Argyriou, C. Broholm, A. Kreyssig, S. Nandi, S. L. Bud'ko, N. Ni, P. C. Canfield, R. J. McQueeney, and A. I. Goldman, Phys. Rev. B {\bf 79}, 060510(R) (2009).

\bibitem{fern} R. M. Fernandes and J. Schmalian, Phys. Rev. B {\bf 82}, 014521 (2010).

\bibitem{liu} C. Liu, A. D. Palczewski, R. S. Dhaka, T. Kondo, R. M. Fernandes, E. D. Mun, H. Hodovanets, A. N. Thaler, J. Schmalian, S. L. Bud'ko, P. C. Canfield, and A. Kaminski, Phys. Rev. B {\bf 84}, 020509(R) (2011).

\bibitem{imai} T. Imai, K. Ahilan, F. L. Ning, T. M. McQueen, and R. J. Cava, Phys. Rev. Lett. {\bf 102}, 177005 (2009).

\bibitem{khas} K. Hashimoto, K. Cho, T. Shibauchi, S. Kasahara, Y. Mizukami, R. Katsumata, Y. Tsuruhara, T. Terashima, H. Ikeda, M. A. Tanatar, H. Kitano, N. Salovich, R. W. Giannetta, P. Walmsley, A. Carrington, R. Prozorov, and Y. Matsuda, Science {\bf 336}, 1554 (2012).

\end{thebibliography}
\end{document}